\documentclass[preprint,10pt,aps, numbers,sort&compress,amsmath,amssymb,amsfonts]{elsarticle}

\usepackage{amssymb}
\usepackage{tikz}
\usepackage{amsmath}
\usepackage{hyperref}
\usepackage{alltt}
\usepackage{slashed}

\usepackage{setspace}
\usepackage{color}
\usepackage{mmacells}

\usepackage{geometry}
\journal{CPC}

\begin{document}

\newcommand{\mhalfo}{\frac{1}{2}}	
\newcommand{\mhalf}[1]{\frac{#1}{2}}
\newcommand{\ka}{\kappa}
\newcommand{\al}{\alpha}
\newcommand{\be}{\beta}
\newcommand{\ga}{\gamma}
\newcommand{\la}{\lambda}
\newcommand{\de}{\delta} 
\newcommand{\vp}[0]{\varphi} 
\newcommand{\vpb}[0]{\bar{\varphi}} 
\newcommand{\equ}[1]{\begin{equation} #1 \end{equation}}
\newcommand{\ba}{\begin{align}}
\newcommand{\ea}{\end{align}}	
\newcommand{\eref}[1]{Eq.~(\ref{#1})}
\newcommand{\fref}[1]{Fig.~\ref{#1}}
\newcommand{\ddotp}[1]{\frac{d^d #1}{(2\pi)^d}}	
\newcommand{\nnnl}{\nonumber\\}	
\newcommand{\G}[1]{\Gamma(#1)}
\newcommand{\nq}{\nu_1}	
\newcommand{\nw}{\nu_2}	
\newcommand{\nd}{\nu_3}	
\newcommand{\dhalf}{\frac{d}{2}} 
\newcommand{\fig}[5]{\begin{figure}[#1]\centering\includegraphics[#5]{#3}\caption{#4}\label{#2}\end{figure}}
\newcommand{\cb}{$\bigstar$} 
\newcommand{\ce}{$\blacksquare$} 
\newcommand{\tw}[1]{\texttt{#1}} 
\newcommand{\beq}{\begin{eqnarray}}
\newcommand{\eeq}{\end{eqnarray}}
\newcommand{\str}{{\rm Tr}}
\newcommand{\hier}{ {\color{red}\rule{1.0\linewidth}{1pt}}}
\newcommand{\colM}[1]{{\color{blue}{#1}}}
\newcommand{\colMT}[1]{{\color{magenta}{#1}}}
\newcommand{\colD}[1]{{\color{red}{#1}}}
\newcommand{\colF}[1]{{\color{green}{#1}}}
\newcommand{\Mathematica}{\textit{Mathematica}}
\newcommand{\DoFun}{\textit{DoFun}}
\newcommand{\DoDSERGE}{\textit{DoDSERGE}}
\newcommand{\DoAE}{\textit{DoAE}}
\newcommand{\DoFR}{\textit{DoFR}}

\begin{frontmatter}



\title{DoFun 3.0: Functional equations in Mathematica}


\author{Markus Q. Huber}
\ead{markus.huber@physik.jlug.de}
\address{Institut f\"ur Theoretische Physik, Justus-Liebig--Universit\"at Giessen, Heinrich-Buff-Ring 16, 35392 Giessen, Germany}
\author{Anton K. Cyrol}
\address{Institut f\"ur Theoretische Physik, Universit\"at Heidelberg, Philosophenweg 16, 69120 Heidelberg,Germany}
\author{Jan M. Pawlowski}
\ead{j.pawlowski@thphys.uni-heidelberg.de}
\address{Institut f\"ur Theoretische Physik, Universit\"at Heidelberg, Philosophenweg 16, 69120 Heidelberg,Germany}
\address{ExtreMe Matter Institute EMMI, GSI, Planckstr. 1, 64291
  Darmstadt, Germany}

\begin{abstract}
  We present version \textit{3.0} of the \textit{Mathematica} package
  \textit{DoFun} for the derivation of functional equations.  In this
  version, the derivation of equations for correlation functions of
  composite operators was added.  In the update, the general workflow
  was slightly modified taking into account experience with the
  previous version.  In addition, various tools were included to
  improve the usage experience and the code was partially restructured
  for easier maintenance.
\end{abstract}

\begin{keyword}
  Dyson-Schwinger equations \sep functional renormalization group
  equations \sep correlation functions \sep quantum field theory \sep
  composite operators


\end{keyword}

\end{frontmatter}

{\bf PROGRAM SUMMARY}

\begin{small}
\medskip 
\noindent
{\em Program Title:} DoFun                                          \\
{\em Version number:} 3.0.0\\
{\em Licensing provisions:} GPLv3
\\
{\em Programming language:}  Mathematica, developed in version 11.3
\\
{\em Operating system:} all on which Mathematica is available (Windows, Unix, MacOS)                                       \\
{\em PACS:} 11.10.-z,03.70.+k,11.15.Tk                                                 \\
\noindent
{\em Nature of problem:} Derivation of functional renormalization group equations, Dyson-Schwinger equations and equations for correlations functions of composite operators in symbolic form which can be translated into algebraic forms.\\
{\em Solution method:} Implementation of algorithms for the derivations of these equations and tools to transform the symbolic to the algebraic form.\\
{\em Unusual features:} The results can be plotted as Feynman diagrams in Mathematica. The output is compatible with the syntax of many other programs and is therefore suitable for further (algebraic) computations.\\
\end{small}

\section{Introduction}

Computer algebra systems are an integral part of particle physics and
physics in general.  Many specialized tools exist and supplement
generic programs like \textit{Mathematica} \cite{Wolfram:2004}.
Especially in perturbative calculations in high-energy physics they
are indispensable, see, e.g.,
\cite{Harlander:1998dq,Baur:2007ub,Luisoni:2016xkv}.  In recent years,
non-perturbative functional methods, see
\cite{Berges:2000ew,Roberts:2000aa,Alkofer:2000wg,Pawlowski:2005xe,
  Fischer:2006ub,Gies:2006wv,Schaefer:2006sr,Binosi:2009qm,
  Braun:2011pp,Maas:2011se,Eichmann:2016yit,Sanchis-Alepuz:2017jjd,Huber:2018ned}
for reviews, have also reached a point where the help of computer
algebra systems is helpful or even mandatory.  To assist in these
cases, a range of dedicated tools was developed
\cite{Alkofer:2008jy,Benedetti:2010nr,Huber:2011qr,Fischbacher:2012ib,Huber:2011xc,Cyrol:2016zqb}.

Here, we present a continuation of that work with a new and extended
version 3 of the program \textit{DoFun} (Derivation Of FUNctional
equations) \cite{Alkofer:2008jy,Huber:2011qr}.  Its purpose is the
derivation of Dyson-Schwinger equations (DSEs), functional
renormalization group equations (RGEs) and - added in this version -
correlation functions of composite operators.  The output can be
arranged in such a way that it is compatible with other programs to
perform traces, like \textit{FormTracer} \cite{Cyrol:2016zqb},
\textit{FORM}
\cite{vanRitbergen:1998pn,Vermaseren:2000nd,Kuipers:2012rf,Kuipers:2013pba,Ruijl:2017dtg},
\textit{HEPMath} \cite{Wiebusch:2014qba} or \textit{FeynCalc}
\cite{Mertig:1990an,Shtabovenko:2016sxi,Shtabovenko:2016whf}, and can be exported to numeric code for numeric calculations \cite{Huber:2011xc}.

In the past, \textit{DoFun} was helpful in several projects using
large systems of equations which are extremely tedious or even
impossible to derive manually.  But even for manually manageable cases
a computer-assisted derivation is very useful.  Examples for the usage
of \textit{DoFun} include Yang-Mills theory and QCD in various gauges
in vacuum, e.g.,
\cite{Alkofer:2008dt,Huber:2009wh,Huber:2009tx,Fister:2010yw,%
  Macher:2011ys,Alkofer:2011pe,Huber:2012zj,Huber:2012kd,%
  Blum:2014gna,Braun:2014ata,Mitter:2014wpa,Huber:2014isa,%
  Cyrol:2014kca,Huber:2014tva,Rennecke:2015eba,%
  Huber:2015ria,Huber:2016tvc,Cyrol:2016tym,Huber:2017txg,Cyrol:2017ewj,Corell:2018yil}
and beyond, e.g.,
\cite{Huber:2016xbs,Cyrol:2017qkl,Contant:2017gtz,Leonhardt:2019fua,Braun:2019aow,Hajizadeh:2019qrj,Contant:2019lwf},
effective models for QCD, e.g.,
\cite{Strodthoff:2016pxx,Pawlowski:2017gxj,Braun:2017srn,%
  Braun:2018bik,Eser:2018jqo,Alkofer:2018guy,Divotgey:2019xea,Leonhardt:2019fua},
asymptotic gravity \cite{Denz:2016qks} and other models
\cite{Janssen:2012pq}.  \textit{DoFun} is often used in combination
with other programs like \textit{Form}
\cite{Ruijl:2017dtg,Kuipers:2013pba,Kuipers:2012rf,Vermaseren:2000nd},
\textit{FormTracer} \cite{Cyrol:2016zqb}, \textit{CrasyDSE}
\cite{Huber:2011xc} or \textit{xPert} \cite{Brizuela:2008ra}.

Since publication of version 2 it became apparent that some aspects of
\textit{DoFun} should be improved to optimize the workflow and in
particular to fully incorporate some cases which were not included in
the original version.  A main consequence of this is a change in the
handling of fields.  Originally the natures of fields were guessed
from the input.  While this makes many use cases simple, it leads to
problems for other cases, e.g., complex scalar fields.  Thus, fields
have to be defined now explicitly which avoids ambiguous situations.

A totally new feature is the derivation of correlation functions for
composite operators.  It relies on a simple identity, but the
calculations can be quite cumbersome as typically many fields are
involved which lead to many loops.  We also added a few new useful
tools, for example, the identification of 1PI diagrams or the
extraction of diagrams of a certain type by name.  The graphical
representation was also modified using now \texttt{Graph[]} instead of
\texttt{GraphPlot[]} which is slightly more versatile.

Finally, we moved the code to a public \textit{git} repository
(\url{https://github.com/markusqh/DoFun}) to make use of a modern
development infrastructure and provide a platform for bug reporting.

In the following we first explain how to install \textit{DoFun} and
get access to the documentation.
Sec.~\ref{sec:installation} also contains a quick start guide.
In Sec.~\ref{sec:funcEqs} we give a short overview of the derivation of
functional equations.  Sec.~\ref{sec:details} contains some additional
details.  We summarize in Sec.~\ref{sec:summary}.  The appendices
contain various summaries of new functions and usage changes from
version 2 to 3.  \textit{For the readers familiar with \emph{DoFun} 2,
  changes they have to consider are listed there as well.}  Readers
who want to get started right away should read the installation
instructions and can then continue with the provided documentation
that includes examples.

\section{Installation and quick start guide}
\label{sec:installation}

\textit{DoFun} was developed in \textit{Mathematica} 11.3 and tested
in \textit{Mathematica} 12. Using it in earlier versions down to 10.0
was not fully tested. It will not work with versions older than version 10.0, since functionality
introduced in that version is used.
In versions 10.0 to 10.2 functionality is limited due to modifications introduced in 10.3.

\textit{DoFun} can be installed with the
installation script from the \textit{git} repository by evaluating
\begin{mmaCell}{Code}
  \mmaDef{Import}["https://raw.githubusercontent.com/markusqh/\
  DoFun/master/DoFun/DoFunInstaller.m"]
\end{mmaCell}
This will put \textit{DoFun} in \textit{Mathematica}'s application folder.\footnote{On a typical Linux system this would be \textit{\textasciitilde/.Mathematica/Applications}.}
Alternatively, one can download it from
\url{https://github.com/markusqh/DoFun/releases} and copy it manually to the
applications folder.

The documentation of \textit{DoFun} is available in
\textit{Mathematica}'s \textit{Documentation Center}: \textit{Add-ons
  and Packages} $\rightarrow$ \textit{DoFun}.  Direct access to the
documentation of a function is also possible via \texttt{??function}
or pressing \textit{F1} when the cursor is inside the function name.

\textit{DoFun} is loaded by evaluating
\begin{mmaCell}{Code}
  \mmaDef{Needs}[\mmaDef{DoFun`}]
\end{mmaCell}

The first step is typically to define the fields which will be used with the newly introduced function \texttt{setFields[]}.
For example, a real bosonic field \texttt{A}, a pair of fermionic fields \texttt{eta} and \texttt{etabar}, and a pair of complex bosonic fields \texttt{phi} and \texttt{phibar} are defined by
\begin{mmaCell}{Code}
 \mmaDef{setFields}[{\mmaDef{A}}, {{\mmaDef{eta}, \mmaDef{etabar}}}, {{\mmaDef{phi}, \mmaDef{phibar}}}};
\end{mmaCell}
As basic ingredient one needs an action.
It can be given in symbolic form as a list of propagators and vertices which themselves are given as lists of fields, e.g.,
\begin{mmaCell}{Code}
 \mmaDef{action} = {{\mmaDef{A}, \mmaDef{A}}, {\mmaDef{eta}, \mmaDef{etabar}}, {{\mmaDef{phi}, \mmaDef{phibar}},
    {\mmaDef{A}, \mmaDef{phibar}, \mmaDef{phi}}, {\mmaDef{A}, \mmaDef{etabar}, \mmaDef{eta}}};
\end{mmaCell}

This is all one needs to start with the derivation of DSEs, flow equations and correlation functions of composite operators using the commands \texttt{doDSE[]}, \texttt{doRGE[]} and \texttt{doCO[]}.
Detailed descriptions for options of these functions can be found in the \textit{Documentation Center}.
For the first two we also refer to the article on \textit{DoFun} 2.0 \cite{Huber:2011qr}.
Below in Sec.~\ref{sec:compOp}, we discuss some aspects specific to \texttt{doCO} and present an example.

The resulting equations can be plotted with the functions \texttt{DSEPlot[]}, \texttt{RGEPlot[]} and \texttt{COPlot[]}.
If desired, one can provide as second argument plot specifications for the fields as illustrated in the example in Sec.~\ref{sec:compOp}.

\section{Derivation of functional equations}
\label{sec:funcEqs}

In this section, the derivation of Dyson-Schwinger and flow equations
is described.  More details can be found in the articles on
\textit{DoDSE} \cite{Alkofer:2008nt} and \textit{DoFun} 2
\cite{Huber:2011qr}, but we reproduce the main steps as a quick
reference.  We also discuss differences in the implementation in
\textit{DoFun} 3.  In Section~\ref{sec:compOp} the equations for
composite operator correlation functions are discussed and exemplified
using the energy-momentum tensor.

\subsection{Basic definitions}

A basic quantity for all derivations is the effective action
$\Gamma[\Phi]$ which depends on the collective field $\Phi$.  An index
encodes the field type, the position or momentum argument and all
indices associated with an internal symmetry group.  Repeated indices
are summed and integrated over if not noted otherwise.  The effective
action is defined via a Legendre transformation:
\begin{align}
 \Gamma[\Phi]&:=\sup_J(-W[J]+J_i \Phi_i ).
\end{align}
The $J_i$'s are the sources for the fields $\Phi_i$.  The generating
functional $W[J]$ is related to the bare action $S[\phi]$ as follows:
\begin{align}
Z[J]=\int D[\phi] e^{-S + \phi_j J_j}=:e^{W[J]}.\label{eq:pathint}
\end{align}
The fields $\phi$ that appear here are the quantum fields which are
related to the average fields $\Phi$ by
\begin{equation}
  \Phi_{i}\equiv \left\langle \phi_{i}\right\rangle _{J}=\frac{
    \delta W}{\delta J_{i}}=Z[J]^{-1}\int D[\phi] \phi_i e^{-S + \phi_j J_j} .
\end{equation}
Setting the sources $J$ to zero leads to the physical expectation
values of the fields $\phi$:
$\Phi_{\rm phys}:=\left\langle \phi_{i}\right\rangle _{J=0}$.

For the effective action a vertex expansion around the physical ground
state is employed:
\begin{equation}
  \Gamma[\Phi]=\sum_{n=0}^{\infty}\,\frac{1}{
    \mathcal{N}^{i_1\ldots i_n}}\sum_{i_1\ldots i_n} \Gamma^{i_1\ldots i_n} 
  (\Phi_{i_1}-\Phi_{i_1,{\rm phys}})\ldots(\Phi_{i_n}-\Phi_{i_n,{\rm phys}}).
\end{equation}
The $\mathcal{N}^{i_1\ldots i_n}$ are symmetry factors.  The physical
$n$-points functions $\Gamma^{i_1\ldots i_n}$ are obtained by
derivatives of the effective action and setting the sources to
zero:\footnote{In contradistinction to \textit{DoFun} 2 we do not
  include a minus sign in the definition of the vertices by default.
  See \ref{sec:signs} for details on signs and how to enable the
  previous behavior for compatibility.}
\begin{subequations}\label{eq:effActions}
\begin{align}
  \Gamma^{ij}:=&\Gamma^{ij}_{J=0}=\frac{\delta ^2 \Gamma[\Phi]}{\de
                 \Phi_{i}\de \Phi_{j}}\Bigg|_{\Phi=\Phi_{\rm phys}},\\
  \Gamma^{i_1\ldots i_n}:=&\Gamma^{i_1\ldots i_n}_{J=0}=\frac{
                            \delta ^{n}\Gamma[\Phi]}{\de \Phi_{i_1}
                            \ldots \de \Phi_{i_n}}\Bigg|_{\Phi=\Phi_{\rm phys}} \label{eq:vertexConvention}.
\end{align}
\end{subequations}
The (field-dependent) propagators are the inverse of the two-point functions:
\begin{align}\label{eq:prop}
 D_J^{ij}:=\frac{\de W[J]}{\de J_i \de J_j}
=\left[\left(\frac{\delta^{2}\Gamma}{\delta\Phi^2}\right)^{-1}\right]^{ij}\,,
\end{align}
Again, the physical propagators are obtained for $J=0$:
\begin{align}
 D^{ij} = D_{J=0}^{ij}.
\end{align}

In the following we will need the derivatives of propagators, fields
and vertices with respect to fields. With the relations 
\begin{align}
  \frac{\delta}{\delta\Phi_{i}}\Gamma^{j_{1}\ldots j_{n}}_{J} & =\frac{\delta\Gamma}{
                                                                \delta\Phi_{i}\delta\Phi_{j_{1}}\ldots\delta\Phi_{j_{n}}}
\end{align}
and 
\begin{align}
  \frac{\delta}{\delta\Phi_{i}}D^{jk}_J                                  &=-\epsilon_i^{jm}\left[\left(\frac{\delta^{2}
                                                                           \Gamma}{\delta\Phi^2}\right)^{-1}\right]^{jm}
                                                                           \left(\frac{\delta^{3}\Gamma}{\delta\Phi_{i}\delta
                                                                           \Phi_{m}\delta\Phi_{n}}\right)\left[
                                                                           \left(\frac{\delta^{2}\Gamma}{\delta\Phi^2}\right)^{-1}\right]^{nk}
\end{align}
we arrive at the simple and complete set of derivative rules 
\begin{subequations}\label{eq:derivatives}
  \begin{align}\label{eq:der1}
\frac{\delta}{\delta\Phi_{i}}\Phi_{j} &=\delta^{ij}\,,\\[2ex]\label{eq:der2}
    \frac{\delta}{\delta\Phi_{i}}\Gamma^{j_{1}\ldots j_{n}}_{J} & =
                                                                  \Gamma^{ij_{1}\ldots j_{n}}_{J}\,,\\[2ex]\label{eq:der3}
   \frac{\delta}{\delta\Phi_{i}}D^{jk}_{J}          &=-\epsilon_i^{jm}D^{jm}_{J}\Gamma^{imn}_{J}D^{nk}_{J}\,. 
\end{align}
\end{subequations}
Note that lowering and raising indices is trivial within the convention
introduced here.  If Grassmann fields are involved, the function
$\epsilon_i^{jm}$ takes care of corresponding signs due to their
anti-commutative nature.  It is defined as
\begin{align}\label{eq:eps}
 \epsilon_i^{jk\ldots}=\left\{
    \begin{array}{ll}
		1 &\quad i \text{ bosonic field}\\[1ex]
		(-1)^{\text{\# Grassmann fields in }jk\ldots} &\quad i \text{ fermionic field}
	\end{array}
	\right .
\end{align}
Note that we use only left-derivatives.\footnote{In \textit{DoFun} 2
left- and right-derivatives were used.}
  
In \textit{DoFun} the effective action $\Gamma[\Phi]$ is generically defined as a list of interactions.
For example, a theory with two fields $\phi_1$ and $\phi_2$ with a quartic interaction is represented by
\begin{mmaCell}{Code}
 {{\mmaDef{phi1}, \mmaDef{phi1}}, {\mmaDef{phi2}, \mmaDef{phi2}}, {\mmaDef{phi1}, \mmaDef{phi1}, \mmaDef{phi2}, \mmaDef{phi2}}}
\end{mmaCell}
At this point no information about the fields is available.
It can be provided by the function \texttt{setFields}.
For the action above we could declare all fields to be bosons by
\begin{mmaCell}{Code}
 \mmaDef{setFields}[{\mmaDef{phi1}, \mmaDef{phi2}}]
\end{mmaCell}
More information is provided in Sec.~\ref{sec:setFields} and the \textit{Documentation Center}.

\subsection{Derivation of Dyson-Schwinger equations}
The master equation is derived from the integral of a total
derivative,
\begin{align}\label{eq:DSE-Z}
  0=&\int D[\phi] \frac{\delta}{\delta \phi_i} e^{-S + \phi_j J_j}
      =\int D[\phi] \left( -\frac{\delta S}{\delta \phi_i}
      + J_i \right) e^{-S + \phi_j J_j}=\left( -\frac{\delta S}{
      \delta \phi'_i}\Bigg\vert_{\phi'_i=\delta/\delta J_i} +J_i \right) Z[J]\,.
\end{align}
Plugging in \eref{eq:pathint}, we can switch to the generating
functional of connected correlation functions, $W[J]$,
\begin{align}
  -\frac{\delta S}{\delta \phi_i}\Bigg\vert_{\phi_i=
  \frac{\delta W[J]}{\delta J_i}
  +\frac{\delta}{\delta J_i}} +J_i=0\,,
\end{align}
where
\begin{align}
  e^{-W[J]}\left(\frac{\delta}{\delta J_i}\right)e^{W[J]}=
  \frac{\delta W[J]}{\delta J_i}+\frac{\delta}{\delta J_i}\,,
\end{align}
was used. Performing a Legendre transformation we obtain the master
equation for 1PI functions:
\begin{align}\label{eq:DSE-master}
  \frac{\delta \Gamma}{\delta \Phi_i}=\frac{\delta S}{\delta
  \phi_i}\Bigg\vert_{\phi_i=\Phi_i+D^{ij}_J  \, \delta/\delta \Phi_j}\,.
\end{align}
By applying further derivatives and setting the sources to zero at the
end, DSEs for any $n$-point function can be obtained.  For more
details we refer to
Refs.~\cite{Alkofer:2008nt,Huber:2011qr,Huber:2018ned} and for a short
description of a graphical derivation to Ref.~\cite{Alkofer:2008jy}.

The implementation of the derivation in \textit{DoFun} is as follows:
\begin{itemize}
 \item Perform the first derivative.
 \item Replace the fields according to \eref{eq:DSE-master}.
 \item Perform additional derivatives.
 \item Set the sources to zero and get the physical propagators and
   vertices.
 \item Get signs from ordering the fermions in a canonical way and the
   $\epsilon$-functions.
 \item Identify equal diagrams.
\end{itemize}
Note that equal diagrams could be identified earlier leading to fewer
intermediate expressions.
However, we use the algorithm described above, because it is simpler.
In case the number of diagrams gets so large that \textit{Mathematica} cannot
handle them anymore, an experienced
user could try as a first simplification to modify this aspect of the algorithm 
in the package code.

The algorithm described above is performed by the function \texttt{doDSE}.
It takes as input an action and the derivatives to perform.
The result is a symbolic expression which can be plotted with \texttt{DSEPlot} or transformed to an algebraic expression with \texttt{getAE}.
For explicit examples we refer to the \textit{Documentation Center} and the article on \textit{DoFun} 2 \cite{Huber:2011qr}.
In Sec.~\ref{sec:compOp} a short example for \texttt{getAE} can be found.

\subsection{Derivation of functional renormalization group equations}

We will follow here the standard derivation of the flow equation for
the so-called effective average action given in
Ref.~\cite{Wetterich:1992yh}.  For flow equations we introduce a
momentum scale $k$ in the bare action $S[\phi]$ via a regulator term.
It serves to integrate out quantum fluctuations in a controlled way:
\begin{align}
 \Delta S_k [\phi] = \frac1{2}\phi_i R^{ij}_k \phi_j\,.
\end{align}
All functionals depend now on $k$.  In the limit $k\rightarrow 0$, the
full effective action is recovered.  The effective average action,
defined by a modified Legendre transformation,
\begin{align}
 \Gamma_k[\Phi]=-W_k[J]+J_i \Phi_i - \frac1{2}\Phi_i R^{ij}_k \Phi_j \label{eq:defavgamma},
\end{align}
is used instead of the standard effective action $\Gamma[\Phi]$.  This
leads to $k$-dependent correlation functions
$\Gamma_k^{i_1\ldots i_n}$. The master equation, which describes the
dependence of the effective average action on the scale $k$, is the Wetterich equation, 
\cite{Wetterich:1992yh}, 
\begin{align}\nonumber 
  \partial_k \Gamma_k[\Phi]=& \frac1{2} \left[\frac{1}{
                              \Gamma_{k}^{(2)}[\Phi] + R_k}\right]^{ji} \partial_k R_k^{ij}\\[1ex]\nonumber 
                              = & \frac1{2}\str \,\frac{1}{
                              \Gamma_{k}^{(2)}[\Phi] + R_k} \partial_k R_k\\[1ex]
                                  =& \frac1{2}\str \, D_{k,J} \partial_k R_k\,,\label{eq:flowEq}
\end{align}
where $\Gamma_{k}^{(2)}[\Phi]$ is the second derivative of the
effective average action.  The trace $\str$ includes a minus sign
for Grassmann fields.  $D_{k,J}$ is the field-dependent propagator
including the regulator term.  Equations for $n$-point functions are
obtained by applying $n$ derivatives to \eref{eq:flowEq} using the
differentiation rules from \eref{eq:derivatives}.

The implementation of the derivation in \textit{DoFun} is as follows:
\begin{itemize}
\item Instead of \eref{eq:flowEq}, the following expression is used to
  minimize the number of diagrams during the derivation (the index $J$
  is suppressed here):
 \begin{align}\label{eq:flowEqLog}
   \partial_t \Gamma_k[\Phi]=& \frac1{2} \str\, \tilde{\partial}_t \ln
                               \left(\Gamma_{k}^{(2)}[\Phi] + R_k \right),
\end{align}
 where $t=\ln (k/\Lambda)$ with $\Lambda$ being a UV cutoff scale.
 The derivative $\tilde{\partial}_t$ only acts on the regulator $R_k$.
 \item The starting expression is
 \begin{align}\label{eq:firstDer}
 \frac1{2}\epsilon_a^{il}\tilde{\partial}_t D^{il}\Gamma^{alj}.
 \end{align}
 The indices $i$ and $j$ are not closed.
 This will be done at the end when also the derivative $\tilde{\partial}_t$ is applied.
 \item Further derivatives are applied with the rules of \eref{eq:derivatives}.
 \item The sources are set to 0 to obtain physical propagators and vertices.
 \item The trace is closed by setting $i=j$.
 If the corresponding propagator belongs to Grassmann fields, a minus sign is added.
 \item The expressions are reorganized in the canonical way, viz., bosons left of anti-Grassmann fields left of Grassmann fields.
 Within these three groups external indices are left of internal ones and the latter are organized by the vertex they connect to.
 This leads to signs from anti-commuting fields and is also required to be able to recognize equal diagrams which are then summed up.
 \item The derivative $\tilde{\partial}_t$ is applied.
\end{itemize}

This algorithm is used by the function \texttt{doRGE} which has the same syntax as \texttt{doDSE}, viz., as input an action and the derivatives are required. The output is a symbolic expression.
Again we refer to the \textit{Documentation Center} and the article on \textit{DoFun} 2 \cite{Huber:2011qr} for specific examples.

\subsection{Correlation functions of composite operators}
\label{sec:compOp}

Any full correlation function can be expressed in terms of dressed
propagators and vertices as \cite{Pawlowski:2005xe}
\begin{align}\label{eq:corrFunc-F}
  \langle F(\phi) \rangle = F\left( \Phi_i+ D_J^{ij}  \, \frac{\delta}{\delta \Phi_j}\right).
\end{align}
For correlation functions of composite operators $O(\phi)$, this leads to
\begin{align}\label{eq:corrFuncCompOp}
  \langle O(\phi(x))O(\phi(y)) \rangle = O\left( \Phi_i(x)+ D_J^{ij}  \, \frac{\delta}{\delta
  \Phi_j}\right)O\left( \Phi_i(y)+ D_J^{ij}  \, \frac{\delta}{\delta \Phi_j}\right)\,,
\end{align}
where the $x$- and $y$-dependence is indicated partially.  The
derivatives to be performed here are similar to the case of DSEs and
we can use the corresponding functions.  For $n$ fields in the
expectation value, up to $n-2$ loops can appear.

For the calculation of \eref{eq:corrFuncCompOp} in \textit{DoFun}, it is convenient to write the composite operator as a general $n$-point function contracted by an auxiliary
function we denote as $C$. $C$ behaves like a vertex and allows using
many functions of \textit{DoFun} in a straightforward way. To
illustrate this, consider the operator
$O_{ij}(x)=\phi_i^a(x) \phi_j^a(x)$. The corresponding two-point
function can be written as (integration and summation over repeated
indices are implied)
\begin{align}
  \langle O_{ij}(x) O_{kl}(y) \rangle = C_{i,j}^{i'a,j'b}(x;x_1,x_2)C_{k,l}^{k'c,l'd}(y;x_3,x_4)
  \langle \phi_{i'}^a(x_1) \phi_{j'}^b(x_2) \phi_{k'}^c(x_3) \phi_{l'}^d(x_4) \rangle
\end{align}
with
\begin{align}
 C_{i,j}^{i'a,j'b}(x;x_1,x_2)=\de^{i'i}\de^{j'j}\de^{ab}\de(x_1-x)\de(x_2-x).
\end{align}

In the following we go step by step through the derivation of the correlation functions of composite operators using a specific example from QCD.
Since in this specific case the result can be simplified further, we show how to realize this with \textit{DoFun}.
The composite operator we will use is the spatial, traceless part of the energy-momentum tensor of Yang-Mills theory:
\begin{align}
  \pi_{ij}(x)=F_{\mu i}^a(x) F_{\,j}^{a,\mu}(x)-\frac{1}{3}\delta_{ij}
  F_{\mu k}^a(x) F^{a,\mu k}(x).
\end{align}
The correlation function we want to calculate is
\begin{align}\label{eq:G}
 G_{\pi\pi}(x,y)=\langle \pi_{ij}(x) \pi_{ij}(y) \rangle,
\end{align}
which, for example, gives access to the shear viscosity via the Kubo
relation \cite{Kubo:1957mj}.  We symbolically write the
energy-momentum tensor as
\begin{align}
 \pi_{ij}=\pi_{ij}^{(2)}+\pi_{ij}^{(3)}+\pi_{ij}^{(4)},
\end{align}
where the numbers in parentheses indicate the number of gluon fields.
$G_{\pi\pi}(x,y)$ can then be split into parts
$G_{\pi\pi}^{(k,l)}(x,y)$ corresponding to pairs of $\pi^{(k)}_{ij}$:
\begin{align}
 G_{\pi\pi}(x,y)=\sum_{k,l=1}^4 \langle \pi_{ij}^{(k)}\pi_{ij}^{(l)} \rangle.
\end{align}
The minimal number of loops appearing in $G_{\pi\pi}^{(k,l)}(x,y)$ is
$\lfloor{(k+l-1)/2\rfloor}$, and the maximal number is $k+l-2$, viz.,
$G_{\pi\pi}(x,y)$ has up to six loops.\footnote{$\lfloor \rfloor$ is
  the floor function.}  We restrict ourselves to two loops here, as
expressions become too long otherwise, but the procedure is the same
for the dropped expressions.  Thus, we only take into account
\begin{align}
  \widetilde{G}_{\pi\pi}(x,y)=\pi_{ij}^{(2,2)}+\pi_{ij}^{(2,3)}+\pi_{ij}^{(3,2)}
  +\pi_{ij}^{(3,3)}+\pi_{ij}^{(2,4)}+\pi_{ij}^{(4,2)}.
\end{align}

The definitions required in \textit{DoFun} are the following. The
composite operator is represented by a field, so we must define it
together with the gluon field:
\begin{mmaCell}{Code}
  \mmaDef{setFields}[{\mmaDef{A},\mmaDef{FF}}]
\end{mmaCell}
We need to define an action which is given as a list of propagators and vertices, each of which is given by a list of fields.
Here we use the Yang-Mills action without ghosts which do not contribute in this case:
\begin{mmaCell}{Code}
  \mmaDef{action}={{\mmaDef{A}, \mmaDef{A}}, {\mmaDef{A}, \mmaDef{A}, \mmaDef{A}}, {\mmaDef{A}, \mmaDef{A}, \mmaDef{A}, \mmaDef{A}}};
\end{mmaCell}
For $\widetilde{G}(x,y)$ we define two auxiliary functions:
\begin{mmaCell}{Code}
\mmaDef{F}[j_] := Module[{j1, j2, j3, j4, j5, j6, j7, j8, j9},
 {\mmaDef{op}[\mmaDef{CO}[{\mmaDef{FF}, j}, {\mmaDef{A}, j1}, {\mmaDef{A}, j2}], {\mmaDef{A}, j1}, {\mmaDef{A}, j2}]/2!,
  \mmaDef{op}[\mmaDef{CO}[{\mmaDef{FF}, j}, {\mmaDef{A}, j3}, {\mmaDef{A}, j4}, {\mmaDef{A}, j5}], {\mmaDef{A}, j3}, {\mmaDef{A}, j4},
    {\mmaDef{A}, j5}]/3!,
  \mmaDef{op}[\mmaDef{CO}[{\mmaDef{FF}, j}, {\mmaDef{A}, j6}, {\mmaDef{A}, j7}, {\mmaDef{A}, j8}, {\mmaDef{A}, j9}], {\mmaDef{A}, j6},
    {\mmaDef{A}, j7}, {\mmaDef{A}, j8}, {\mmaDef{A}, j9}]/4!}]

\mmaDef{pi}[i_, j_, k_, l_] := \mmaDef{op}[\mmaDef{F}[i][[k - 1]], \mmaDef{F}[j][[l - 1]]]
\end{mmaCell}
The second quantity is the combination of the parts with $k$ and $l$ gluon legs.
$\widetilde{G}(x,y)$ can then be written as
\begin{mmaCell}{Code}
 \mmaDef{G1} = \mmaDef{pi}[i, j, 2, 2] + \mmaDef{pi}[i, j, 2, 3] + \mmaDef{pi}[i, j, 3, 2] + 
   \mmaDef{pi}[i, j, 3, 3] + \mmaDef{pi}[i, j, 2, 4] + \mmaDef{pi}[i, j, 4, 2];
\end{mmaCell}
The next steps are the replacements of the fields according to \eref{eq:corrFunc-F} and setting the sources to zero:
\begin{mmaCell}{Code}
 \mmaDef{G2} = \mmaDef{replaceFields}[\mmaDef{G1}];
 \mmaDef{G3} = \mmaDef{setSourcesZero}[\mmaDef{G2}, \mmaDef{action}, {}];
\end{mmaCell}
As mentioned above, we only take diagrams up to two loops:
\begin{mmaCell}{Code}
 \mmaDef{G4} = Select[\mmaDef{G3}, \mmaDef{getLoopNumber}[#] <= 2 &];
\end{mmaCell}
The steps from \texttt{G1} to \texttt{G4} can be obtained directly with the function \texttt{doCO}:
\begin{mmaCell}{Code}
\mmaDef{G4} = \mmaDef{doCO}[\mmaDef{action}, \mmaDef{G1}, \mmaDef{getLoopNumber}[#] <= 2 &];
\end{mmaCell}
The last argument used is optional and selects here the terms with one and two loops.

\begin{figure}[tb]
 \begin{center}
  \includegraphics[width=\textwidth]{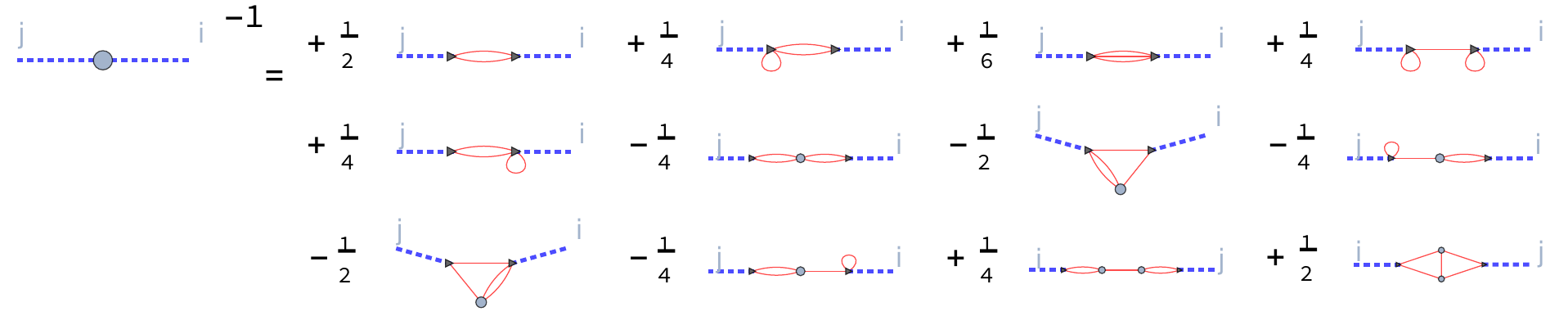}
 \end{center}
 \caption{The correlation function \eref{eq:G} up to two loops.  Red,
   continuous lines are gluons, the triangle represents the composite
   operator $\pi_{ij}$ with the external leg plotted blue and dotted.}
 \label{fig:G_res1}
\end{figure}

$G4$ contains connected and disconnected diagrams. Since the relevant
quantity is actually the expectation value of the commutator of the
composite operator, the latter will finally vanish, and we drop them here.
Of the originally 72 diagrams, many of which are identical, though,
now 63 remain:
\begin{mmaCell}{Code}
\mmaDef{GConn} = \mmaDef{getConnected}[\mmaDef{G4}];
\end{mmaCell}
We sum up those graphs:
\begin{mmaCell}{Code}
 \mmaDef{GConnId} = 
  \mmaDef{identifyGraphs}[\mmaDef{GConn}, {{\mmaDef{FF}, i}, {\mmaDef{FF}, j}}];
\end{mmaCell}
Using the plot styles
\begin{mmaCell}{Code}
\mmaDef{fieldRules} = {{\mmaDef{A}, Red}, {\mmaDef{FF}, \mmaDef{Thick}, Orange}};
\end{mmaCell}
we can plot the result:
\begin{mmaCell}{Code}
 \mmaDef{COPlot}[\mmaDef{GConnId}, \mmaDef{fieldRules}]
\end{mmaCell}
It is depicted in \fref{fig:G_res1}.  We see three types of diagrams
which are not 1PI.  They all vanish, because either the combination of
a three-gluon vertex and a two-gluon-leg part of $\pi_{ij}$ yields
zero due to the color structure \cite{Haas:2014th} or because the
combination of a gluon propagator and the three-gluon-leg part of
$\pi_{ij}$ yields zero.

\begin{figure}[tb] 
 \begin{center}
  \includegraphics[width=\textwidth]{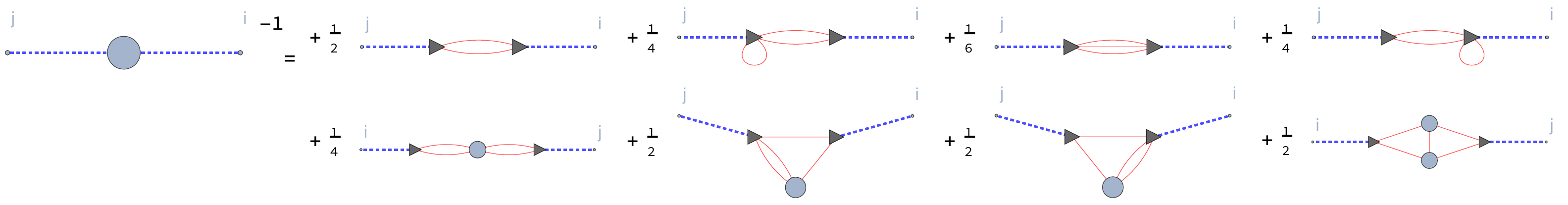}
 \end{center}
 \caption{The correlation function \eref{eq:G} up to two loops with
   some diagrams discarded that vanish due to color contractions.}
 \label{fig:G_final}
\end{figure}

We thus continue only with the 1PI part:
\begin{mmaCell}{Code}
 \mmaDef{G} = \mmaDef{get1PI}[\mmaDef{GConnId}]; 
\end{mmaCell}
This is the final result in symbolic form.
A graphical representation is depicted in \fref{fig:G_final}. 

We can use \textit{DoFun} to convert this to an algebraic expression.
However, we will only explain schematically how to do this, as the resulting expressions are very long and not very instructive.
As a first step, we define which indices the fields have:
\begin{mmaCell}{Code}
  \mmaDef{defineFieldsSpecific}[{\mmaDef{A}[mom, col, lor],
    \mmaDef{FF}[mom, lors, lors]}]
\end{mmaCell}
This assigns the gluon field $A$ a momentum, a color and a Lorentz
index and the energy-momentum tensor a momentum and two spatial
Lorentz indices.\footnote{The names of the indices are arbitrary and
do not have any special meaning within \textit{DoFun}.} 
In the next step, we would need to define the Feynman rules for the gluon propagator,
the three- and four-gluon vertices and the components of the
energy-momentum tensor.
Once this is done, one can transform the symbolic to the algebraic expressions with
\begin{mmaCell}{Code}
 \mmaDef{getAE}[\mmaDef{G}, {{\mmaDef{FF}, i, p, k, l}, {\mmaDef{FF}, j, -p, m, n}}]
\end{mmaCell}
The arguments in the lists assign, for example, the external field
$FF$ with the index $i$ the momentum $p$ and the spatial Lorentz
indices $k$ and $l$.
This is the final expression.
To continue from here, one can, for example, perform traces or export the expression to a numeric program.

\begin{figure}[tb]
 \begin{center}
  \includegraphics[width=\textwidth]{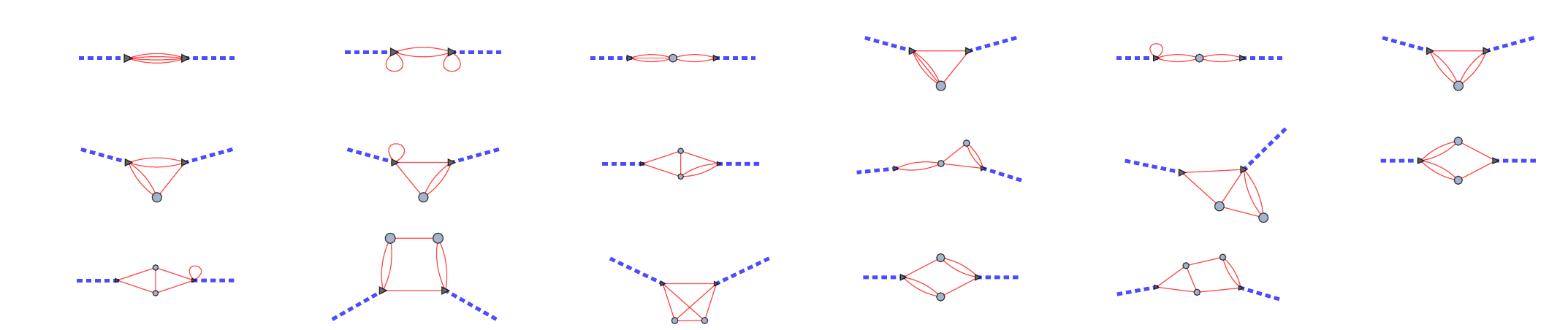}
 \end{center}
 \caption{Examples for three-loop contributions to $G(x,y)$.}
 \label{fig:G_L3}
\end{figure}

Finally, we show some three-loop diagrams as an example of higher
contributions to \eref{eq:G} in \fref{fig:G_L3}.

\section{Some details}
\label{sec:details}

This section collects some more detailed information about certain
aspects of \textit{DoFun}.

\subsection{Structure of \textit{DoFun}}

The main part of \textit{DoFun} is contained in three package files:
\begin{itemize}
\item \textit{DoDSERGE.m}: This package contains the functions for
  deriving DSEs, flow equations and composite operator correlation
  functions.
\item \textit{DoFR.m}: This package provides tools to derive Feynman
  rules from a given action.  Changes in version \textit{3} are minor
  and it should work with old programs without problems.  Only signs
  for fermions should be checked because all derivatives are
  left-derivatives now.
\item \textit{DoAE.m}: This package contains functions to transform
  symbolic to algebraic expressions.  The functionality is basically
  identical to version \textit{2} with the exception that composite
  operators were added.  It should work with old programs without
  problems.
\end{itemize}

Additional files/directories are the following:
\begin{itemize}
\item \textit{Kernel/init.m} is called for loading \textit{DoFun} and
  automatically checks for updates.
 \item \textit{DoFunInstall.m} can be called for installing \textit{DoFun} on a computer.
 It copies all required files to the appropriate places.
\item \textit{Documentation} contains the documentation which can be
  accessed in the \textit{Documentation Center} via \textit{Add-ons
    and Packages} $\rightarrow$ \textit{DoFun}.
\end{itemize}

\subsection{Fields}
\label{sec:fields}

All fields are assigned a specific type out of the following list:
\texttt{boson}, \texttt{fermion}, \texttt{antiFermion},
\texttt{complex} and \texttt{antiComplex}.  The field type is obtained
with \texttt{fieldType[]}.  To assign these types, the function
\texttt{setFields[]} is used.  It also takes care of setting various
other properties of fields, e.g., if they are commuting or
anti-commuting, which can be checked with \texttt{cFieldQ[]} and
\texttt{grassmannQ[]}.  Fields know about their anti-fields which can
be determined with \texttt{antiField[]}.

\subsection{Treatment of anti-commuting fields}

Signs from the anti-commuting nature of fermions are taken into
account by the function defined in \eref{eq:eps}.  It is called
\texttt{sf[]} and automatically simplifies in many cases.  It is
inserted whenever needed and only at the end, when all single fields
of the superfield are written out, the signs are determined with the
function \texttt{getSigns[]}.

The notation for vertices uses the following convention for the places of indices:
\begin{align}\label{eq:Gamma_psibpsibpsipsi}
  \Gamma^{\bar \psi \bar \psi \psi\psi}_{ijkl} = \frac{\delta^4 \Gamma}{ \delta\bar\psi_i
  \delta\bar \psi_j \delta\psi_k \delta\psi_l}
\end{align}
This entails that fields in the derivation functions are put in
inverse order, because they are applied from left to right.  For
example, the DSE corresponding to \eref{eq:Gamma_psibpsibpsipsi} is
derived by
\begin{mmaCell}{Code}
  \mmaDef{doDSE}[\mmaDef{action}, {{\mmaDef{psi}, l}, {\mmaDef{psi}, k}, {\mmaDef{psibar}, j}, {\mmaDef{psibar}, i}}]
\end{mmaCell}

\section{Summary}
\label{sec:summary}

Calculations in quantum field theory can easily become too tedious to
perform them manually.  Automatizing them not only alleviates the
calculations but is in some cases mandatory.  The software
\textit{DoFun} we presented here is a tool for functional calculations
which can profit from such an automatization.  It can derive various
functional equations from a given action and yields results in a form
suitable for further manipulations by other programs.  The present
version 3 simplifies some aspects, introduces new tools and adds
composite operators to the tool box.

\section{Acknowledgments}

We thank Jens Braun, Tobias Denz, Marc Leonhardt, Mario Mitter,
Coralie Schneider, Nils Strodthoff and Nicolas Wink for useful
discussions and input on how to improve \textit{DoFun}.  Funding by
the FWF (Austrian Science Fund) under Contract No. P27380-N27 is
gratefully acknowledged. The work is also supported by EMMI, the BMBF
grant 05P12VHCTG, and is part of and supported by the DFG
Collaborative Research Centre SFB 1225 (ISOQUANT) as well as by the
DFG under Germany's Excellence Strategy EXC- 2181/1 - 390900948 (the
Heidelberg Excellence Cluster STRUCTURES).

\appendix

\section{Usage changes from \textit{DoFun} 2}

This section lists differences between \textit{DoFun} 2 and 3 and is
intended for users already familiar with the former.  New users can
skip this section.

\subsection{Quick fact sheet on how to update to \textit{DoFun} 3}

\begin{tabular}{|l|p{10cm}|}
  \hline
  Task & Description\\
  \hline\hline
  Field definitions & Fields need to be explicitly declared before any calculations with \texttt{setFields[]}.\\
  \hline
  Sign conventions & Vertices are defined now as the positive derivative of the effective action which leads to additional minus signs in diagrams. The old behavior can be restored by setting \texttt{\$signConvention=1} before any derivations.\\
  \hline 
  Derivatives & All derivatives are now left-derivatives. This may affect signs of fermionic vertices.\\
  \hline
  Algebraic expressions & The order of the resulting field arguments of propagators and vertices might be different now.
  When converting the symbolic into algebraic expressions, the definitions of propagators and vertices might thus need to be adapted.\\
  \hline
\end{tabular}

\subsection{Types of fields and \texttt{setFields[]}}
\label{sec:setFields}

Fields have an explicit type now.  They can be (real) bosons, complex
fields or fermions.  To have a clear connection between fields and
their anti-fields, the latter have the types anti-complex field and
anti-fermionic, respectively.  The field type can be obtained with
\texttt{fieldType[]}.  Before doing any calculations, the field type
has to be declared with \texttt{setFields[]}.  It also sets other
properties like if they are commuting or
anti-commuting, which can be checked with \texttt{cFieldQ[]} or \texttt{grassmannQ[]}, respectively.

\subsection{Sign conventions}
\label{sec:signs}

\begin{itemize}
 \item The definition of vertices changed compared to \textit{DoFun} 2.
 They are defined now as
 \begin{align}
   \Gamma^{i_1\ldots i_n}:=&\Gamma^{i_1\ldots i_n}_{J=0}=\frac{\delta ^{n}
                             \Gamma[\Phi]}{\de \Phi_{i_1}\ldots \de \Phi_{i_n}}\Bigg|_{\Phi=\Phi_{\rm phys}}
                             \label{eq:vertexConventionApp}.
\end{align}
  This is determined by the default value of \texttt{\$signConvention=-1}.
  It can be reset to $1$ to recover the old behavior.
 \item For Grassmann-valued fields only left-derivatives are used.
  This can lead to different signs compared to using left- and right-derivatives.
\end{itemize}

\section{New functions}

Various new functions were added.
The ones accessible for the user are listed below:

\begin{itemize}
 \item \texttt{setFields[]}: Defines properties of fields, see Sec.~\ref{sec:setFields}.
 \item \texttt{doCO[]}: Derive equation for the correlation function of a composite operator.
 \item \texttt{getVertexNumbers[]}, \texttt{getDiagramType[]},
   \texttt{extractDiagramType[]} and \texttt{groupDiagrams[]}: Tools
   to classify and extract diagrams.  Known diagram types are stored
   in \texttt{diagramTypes}, which can be extended by definitions of
   the user.  For example, all diagrams of triangle type can be
   extracted from the expression \texttt{exp} by
\begin{verbatim}
 extractDiagramType[exp, "triangle"]
\end{verbatim}
   Currently the following diagrams are defined: oneLoop, tadpole,
   sunset, squint, triangle3, swordfish3, box, triangle4, swordfish4,
   fivePoint4
 \item \texttt{sortCanonical[]}: Puts fields in a canonical order as required for diagram identification. 
 Replaces the function \texttt{orderFermions[]}.
\item \texttt{getSigns[]}, \texttt{sf[]}: Signs from permuting fields
  are determined with \texttt{getSigns[]} using the information stored
  in the auxiliary function \texttt{sf[]}.
\item \texttt{getConnected[]},
  \texttt{getDisconnected[]}, \texttt{connectedQ[]}, \texttt{disconnectedQ[]}: Extracts (dis)connected diagrams.
 \item \texttt{getNon1PI[]}, \texttt{get1PI[]}, \texttt{onePIQ[]}: Extracts 1PI/non-1PI diagrams.
\end{itemize}

\section{Limitations and disclaimer}

\textit{DoFun} was carefully tested.  Nevertheless we cannot guarantee
that the program is free of flaws.  However, we encourage everybody
who finds a bug to report it on \textit{GitHub} via
\url{https://github.com/markusqh/DoFun/issues}.

At the moment of publication of this article, we know of the following limitations with respect to diagram identification and plotting expressions.
The symbolic and algebraic results, though, are correct.
\begin{itemize}
\item Identification of diagrams only works up to two-loop.  It should
  be noted that graph identification is a non-trivial problem in
  general.  We decided not to put too much effort into this, because
  in relevant cases the identifications can still be done manually.
  In practice, we think the only relevant case are equations of
  composite operators. Identification can also fail when mixed propagators appear.
\item \textit{Mathematica} 12.0.0 introduced some changes in plotting
  graphs which either introduced a bug or removed a certain feature on purpose which was used in \textit{DoFun} 2.0.  Thus, it may cause problems to plot diagrams
  with more than two propagators connecting the same vertices.  A
  warning message is printed if this happens and the styles of the
  propagators may be incorrect.
\end{itemize}

\bibliographystyle{model1a-num-names}
\bibliography{DoFun3_literature}

\end{document}